\documentclass[aps,reprint,floatfix,twocolumn,prl]{revtex4-1}
\usepackage{xcolor}
\usepackage{graphicx}
\usepackage[hidelinks]{hyperref}
\usepackage{amssymb}
\usepackage{amsmath}

\newcommand{\change}[1]{\textcolor{blue}{#1}}

\newcommand{\be}{\begin{equation}}
\newcommand{\ee}{\end{equation}}
\renewcommand{\section}[1]{\noindent\textit{#1}.}
\begin{document}
\title{Hund nodal line semimetals: The case of twisted magnetic phase in the double-exchange model}

\author{R. Matthias Geilhufe$^1$}
\author{Francisco Guinea$^{2,3}$}
\author{Vladimir Juri\v ci\' c$^1$}
\affiliation{$^1$Nordita,  KTH Royal Institute of Technology and Stockholm University, Roslagstullsbacken 23,  10691 Stockholm,  Sweden\\$^2$Imdea Nanoscience, Faraday 9, 28015 Madrid, Spain \\$^3$ School of Physics and Astronomy. University of Manchester, Manchester M13 9PY}

\date{\today}

\begin{abstract}
We propose a class of topological metals, which we dub \emph{Hund nodal line semimetals}, arising from the strong Coulomb interaction encoded in the Hund's coupling between itinerant electrons and localized spins. We here consider a particular twisted spin configuration, which is realized in the double exchange model which describes the manganite oxides. The resulting effective tetragonal lattice of electrons with hoppings tied to the local spin features an antiunitary  \emph{non-symmorphic} symmetry that in turn, together with another non-symmorphic but unitary glide mirror symmetry protects crossings of a double pair of bands along a high-symmetry line on the Brillouin zone boundary. We also discuss the stability of Hund nodal line semimetal with respect to  symmetry breaking arising from various perturbations of the twisted phase. Our results motivate further studies of other realizations of this state of matter, for instance in different spin backgrounds, properties of its drumhead surface states, as well as its stability to  disorder and interactions among the itinerant electrons.
\end{abstract}
\maketitle
\section{Introduction}
Weyl semimetals, paradigmatic representatives of topological metals,  have recently attracted considerable attention in both theoretical and experimental condensed matter communities  due to their intriguing properties, such as unusual Fermi arc surface states and chiral anomaly \cite{review1,armitage-RMP}. Due to either broken time-reversal or inversion symmetry, they host chiral Weyl fermions at isolated points in the Brillouin zone (BZ). In contrast, Dirac semimetals host four-fold degenerate Dirac points, in the presence of time-reversal and inversion symmetry~\cite{murakami2007,yang2014}.

Band crossings can also occur on a manifold of higher dimensionality, such as a line, in the BZ, yielding nodal line semimetals (NLSMs)~\cite{burkov1}. NLSMs feature localized drumhead surface states forming flat electronic bands thereby providing a platform for a possible realization of interaction driven states, such as superconductors, magnetic phases, charge density waves, and anomalous Hall states~\cite{volovik1,volovik2,bitan,uchoa-2}. As their point-like analogs, NLSMs can be of a Dirac or a Weyl type, depending on the degeneracy of the bands crossing at the nodal line. Topology and various symmetries protect NLSMs, including a combination of inversion and time-reversal~\cite{Ben2} and non-symmorphic lattice symmetries~\cite{CFang-2015,Matthias1}, both with and without spin-orbit coupling~\cite{Xi-Review-TNLSM2016}. Recently, NLSMs have been theoretically proposed and experimentally realized in various materials, for instance, ZrSiS~\cite{neupane1,Schoop2016}, HfSiS~\cite{Chen-1,takane}, HfC~\cite{RuiYu3}, Cu$_3$PdN\cite{RuiYu2}, CaP$_3$~\cite{RuiYu1}, TlTaSe$_2$~\cite{bian-1}, and graphene-like three-dimensional systems~\cite{uchoa-1,Weng-PRB2015}; see also Ref. ~\cite{Yang-AdvPhysX}.

 The emergence of topological states in gapless systems in conjunction with the non-magnetic lattice symmetries is by now rather well understood ~\cite{Slager2013,schnyder-RMP,Bradlyn2016,Bradlyn2017,Slager2017,Po2017}. In contrast, topological metallic states in magnetic lattices have just begun to be explored~\cite{BenW-2017,Cuoco-2017,Rauch-2017,wang2017,Yu2017,Vishwanath1}. In that respect, the questions regarding possible topological phases emerging out of magnetic lattices with non-collinear spins and the specific mechanisms are still pertinent, and we address these in the present paper.

We study magnetic phases in a broad class of materials, such as manganites or Colossal Magneto Resistance systems (CMR's)~\cite{CVM99}. These materials exhibit a large number of magnetic phases~\cite{alonso2001,WK55,tikkanen2016}, which can be tuned by doping or a magnetic field. Most of them are metals well described by the double exchange model~\cite{Z51,AH55,G60}.
For manganites such as La$_{1-x}$Ca$_x$MnO$_3$, the oxidation state of Mn fluctuates between Mn$^{3+}$ and Mn$^{4+}$ as a function of doping.
Additionally, these states give rise to strongly localized spins due to the alignment of the three $t_{2g}$ orbitals in the crystal-field split $d$ band of Mn (in a cubic lattice) exhibiting a total spin of $S = 3/2$. Due to a strong intra-atomic Hund's coupling, the carriers' spins are aligned parallel to the core spin, such that the magnetic phase influences the hopping of the delocalized carriers between neighboring Mn sites. The rich variety of magnetic phases
was initially assigned to canting of the core spins. It was later shown that electronic phase separation is also likely~\cite{Yetal98,AG98}, which explains the observed hysteretic behavior of many CMR systems.

Here, we show that the properties of double exchange materials allow us to define a class of NLSMs, which we dub \emph{Hund nodal line semimetals}, arising from a strong \emph{Hund's} coupling between itinerant and localized electrons forming non-collinear spin configuration~\cite{footnote}.
Specifically, we consider the twisted magnetic phase~\cite{alonso2001}, which is realized in the double exchange model as discussed in the context of manganites~\cite{Dagotto-2001,Salamon-RMP2001,footnote1}.
Strong Hund's coupling leads to the emergence of an effective spin lattice, which, as we show, features an antiunitary \emph{non-symmorphic} symmetry, that, together with the unitary glide mirror symmetry, protects the Hund nodal line semimetallic state.
To corroborate this mechanism, we consider various perturbations of the twisted phase, and explicitly show  that the nodal line is stable as long as the resulting Hamiltonian on the BZ boundary respects the protecting symmetries.

\section{The Model} The double exchange model is characterized by a strong Hund's coupling between the localized spin ${\bf S}_i$ and the itinerant electron spin ${\bf s}_i$ at site $i$, with ${\bf s}_i=\sum_{\alpha,\beta}c_{i,\alpha}^\dagger {\boldsymbol \xi}_{\alpha\beta}c_{i,\beta}$. Here, $c_{i,\alpha}$ is the annihilation operator for an electron with spin projection $\alpha=\uparrow,\downarrow$, and ${\boldsymbol\xi}$ are Pauli matrices. Projecting out the electrons' spin component antiparallel to the localized spins leads to an effective
tight-binding model for the itinerant electrons but now with the spin parallel to ${\bf S}_i$~ \cite{Nolting-book},
\begin{equation}\label{TB}
H=t\sum_{ij}[\langle\theta_i\phi_i|\theta_j\phi_j\rangle c_{i}^\dagger c_j+h.c.].
\end{equation}
The hopping elements are given in terms of the overlap $\langle\theta_i\phi_i|\theta_j\phi_j\rangle$ between the localized spins
and $c_i$ is the annihilation operator for an electron (after the projection) at site $i$.
Here, the localized spin ${\bf S}$ is described as a classical three-dimensional unit vector, $|{\bf S}|=1$,
with spherical angles $(\theta_,\phi)$. In addition, the localized spins are coupled through the super-exchange interaction, which together with the effective hopping Hamiltonian in Eq.~(\ref{TB}) defines the double-exchange model.

\section{Twisted phase}
\begin{figure}
\includegraphics[]{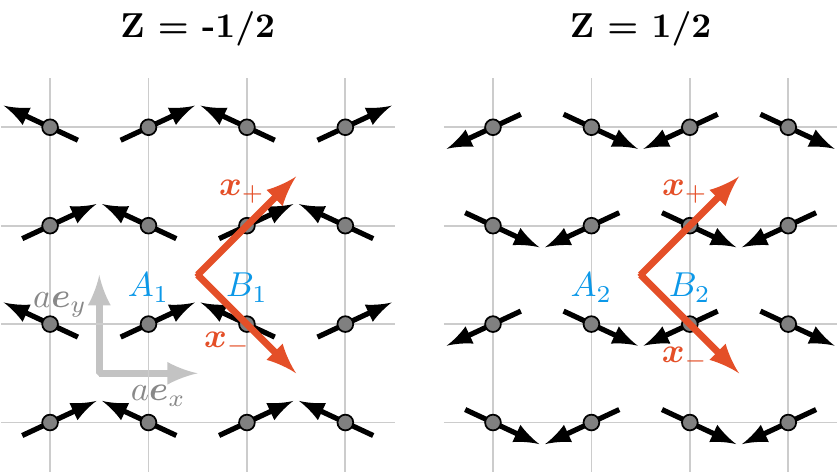}
\caption{Spin configuration in the two sublattices.\label{spin-conf}}
\end{figure}

\begin{figure}[b!]
\centering
\includegraphics[width=0.49\textwidth]{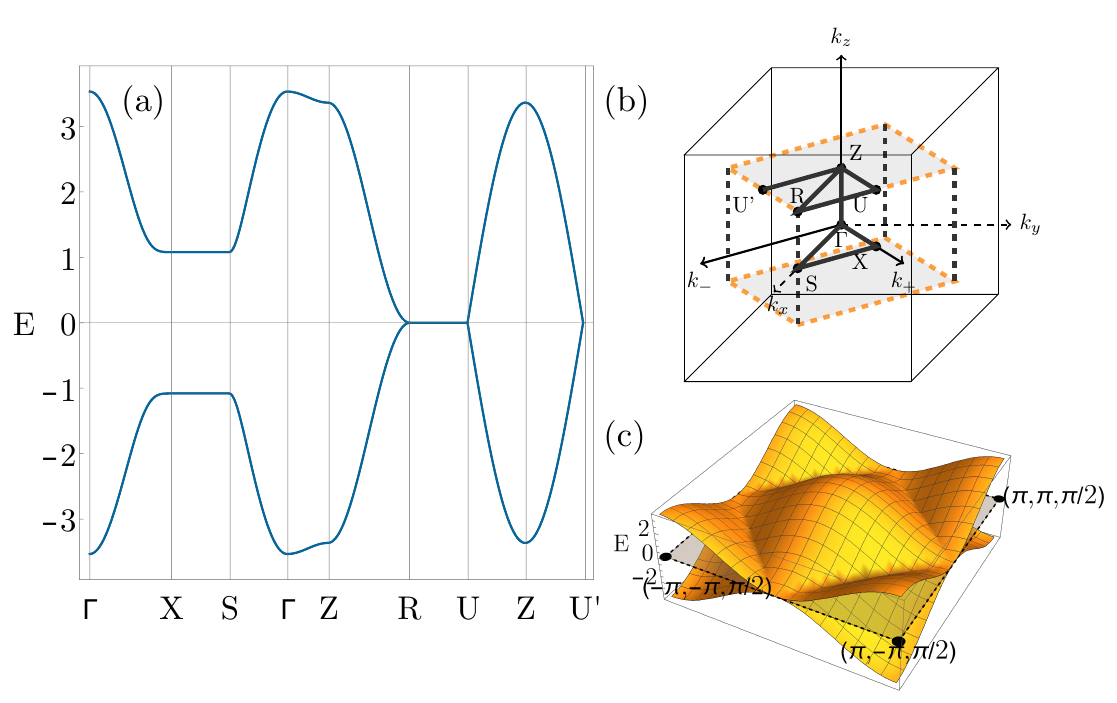}
 \caption{Electronic structure of the pristine model with $\phi=2$ and $a=t=1$.
 (a) Band structure calculated along a high symmetry path. (b) Energy bands forming a nodal line for $z=\pi/2$.  (c) Position and shape of the nodal line in the Brillouin zone together with high-symmetry points.}
 \label{fig:LNBZ}
 \end{figure}

We consider the itinerant electrons hopping in the background of the localized spins in the twisted phase given by~\cite{alonso2001}
\begin{equation}
{\bf S}_i=\cos\phi(-1)^{x+y} {\bf e}_x+\sin\phi(-1)^{z} {\bf e}_y,
\label{spin:eq}
\end{equation}
where $(x,y,z)$ are the Cartesian coordinates of the site $i$ of a cubic lattice in units of the lattice spacing ($a$) and ${\bf e}_j$, $j=x,y,z$, is the unit vector
in the lattice direction $j$. Since the spin is confined to the $x-y$ plane, it can be parametrized by a polar angle $\phi$. This spin configuration spans two sublattices in the $x-y$ plane (Fig. \ref{spin-conf}). Additionally, the two adjacent planes feature inequivalent spin configurations.
We therefore construct a unit cell with four sites labeled as $\alpha_k$, $\alpha=a, b$ is the sublattice index in the $x-y$ plane and $k=1,2$ denotes the two inequivalent adjacent $x-y$ planes. The lattice translations are generated by ${\bf x}_\pm=a\sqrt{2}({\bf e}_x\pm{\bf e}_y)\equiv \tilde{a}({\bf e}_x\pm{\bf e}_y)$ and ${\bf z}=2a{\bf e}_z$. The Brillouin zone is diamond shaped in the $x-y$ plane: $-\frac{\pi}{{\tilde a}}\leq k_\pm\leq \frac{\pi}{{\tilde a}},\,-\frac{\pi}{{2 a}}\leq k_z\leq \frac{\pi}{2{a}}$, with $k_\pm =k_x\pm k_y$. The four distinct sites exhibit the four spin states $|\pm\phi\rangle$ and $|\pm\phi+\pi\rangle$, given by
\begin{align}
|\pm\phi\rangle &=\frac{1}{\sqrt{2}}\left(
    \begin{array}{c}
      1 \\
      e^{\pm i\phi}\\
    \end{array}
  \right), |\pm\phi + \pi \rangle=\frac{i}{\sqrt{2}}\left(
    \begin{array}{c}
      1 \\
      -e^{\pm i\phi}\\
    \end{array}
  \right).
\end{align}
Considering only nearest neighbor hoppings, the tight-binding Hamiltonian in Eq.~(\ref{TB}) can be written as $H_{tw}^{NN}=\sum_{{\bf k}}\psi^\dagger_{\bf k} H_{tw}^{NN}({\bf k})\psi_{{\bf k}}$ \cite{SM}
\begin{align}\label{eq:TBmomentumspace}
H_{tw}^{NN}({\bf k})&=(\cos k_x+\cos k_y)\left[-(1-\cos2\phi)\sigma_0\otimes\tau_2 \right. \nonumber\\ &- \left.\sin2\phi\, \sigma_3\otimes\tau_1\right]
+\cos k_z\left[(1+\cos2\phi)\sigma_1\otimes\tau_0 \right. \nonumber\\ &+\left. \sin2\phi\, \sigma_2\otimes \tau_3\right],
\end{align}
in  the spinor basis  $\psi_{\bf k}=(a_{1,{\bf k}},b_{1,{\bf k}},a_{2,{\bf k}},b_{2,{\bf k}})^\intercal$ with $\alpha_{s,i}$ as the annihilation at a site $i$ belonging to a sublattice $\alpha=a,b$ in the one of the two inequivalent $x-y$ planes $s=1,2$.
The Pauli matrices $\sigma_i$ and $\tau_i$ act in the $(1,2)$ and $(a,b)$ spaces, while $\sigma_0$ and $\tau_0$ are $2\times2$ unity matrices; the lattice spacing of the original cubic lattice, $a$, and the overall energy scale, $t$, are both set to unity. This Hamiltonian yields twofold degenerate valence and conduction bands
\begin{equation}
E_{\pm,{\bf k}}=\pm2\sqrt{(\cos k_x+\cos k_y)^2\sin^2\phi+\cos^2k_z\cos^2\phi}.
\end{equation}
The obtained band structure features diamond shaped nodal lines at $k_x\pm k_y=\pm \pi$ in each of the Brillouin zone boundary planes  $k_z=\pm\pi/2$ (Fig.~\ref{fig:LNBZ}).
Importantly, at low energies the NLSM exhibits linear  dispersion in the directions perpendicular to it, with a momentum-dependent Fermi velocity in the $x-y$ plane $\sim \sin k_x$, with $k_x$ along the nodal line.

\section{Symmetry group of the system and protection of the line nodes}
The four sites in the unit cell are associated to the spin states $|\phi\rangle$, $|-\phi\rangle$, $|\phi+\pi\rangle$, and $|\pi-\phi\rangle$. Starting from $|\phi\rangle$, we construct three symmetry elements that realize the mapping to the remaining three sites (Table \ref{sym:elements}).

First, the complex conjugation ${\rm K}$ maps $\phi\rightarrow-\phi$. In the basis $\psi_{\bf{k}}$ this operation corresponds to the permutations $a_1\leftrightarrow a_2$ and $b_1\leftrightarrow b_2$, which can be expressed as $(\sigma_2\otimes\tau_0){K}$. We therefore conclude that the combination of complex conjugation accompanied by a fractional shift in the unit cell along the $z$-axis is a symmetry of the Hamiltonian and denote it by $\left\{{\rm K}|0,0,1/2\right\}$.

 \begin{table*}[t!]
\begin{tabular}{lllll}
 \hline\hline
Symmetry element & action on $|\phi\rangle$ & action on $(x_+,y_+,z)$ & action on $(k_+,k_-,k_z)$ & action on site-basis \\
   \hline
   $\left\{{\rm K}|0,0,1/2\right\}$ & $|-\phi\rangle$ & $(x_+,x_-,z+1/2)$ & $(-k_+,-k_-,-k_z)$ & $(\sigma_2\otimes\tau_0){K}$ \\
   $\left\{{\rm IC_{2z}}|1/2,-1/2,0\right\}$ & $|\phi+\pi\rangle$ & $(x_++1/2,x_--1/2,-z)$ & $(k_+,k_-,-k_z)$ & $i(\sigma_1\otimes\tau_2)$ \\
   $\left\{{\rm KIC_{2z}}|1/2,-1/2,1/2\right\}$ & $|\pi-\phi\rangle$ & $(x_++1/2,x_--1/2,-z+1/2)$ & $(-k_+,-k_-,k_z)$ & $-i(\sigma_3\otimes\tau_2){K}$ \\
\hline\hline
\end{tabular}
\caption{Symmetry operations and their representation in  the spin basis, in the real and momentum space, as well as in the pseudo-spinor basis $\psi_{\bf k}$ defined by the unit cell in the twisted phase. \label{sym:elements} }
\end{table*}

Second, consider the twofold rotation about the $z$-axis, ${\rm C}_{2z}:(x_+,x_-,z)\rightarrow(-x_+,-x_-,z)$, which operates in spin space as $\exp(i \xi_3 \pi/2 )=i\xi_3$. Therefore, under ${\rm C}_{2z}$,  the spin state  $|\pm\phi\rangle\rightarrow |\pi\pm\phi\rangle$, which corresponds to the permutation $a1\leftrightarrow b2, a2\leftrightarrow b1$. This implies that the combined operation of ${\rm C}_{2z}$, inversion ${\rm I}$ (swapping the planes $1\leftrightarrow 2$) and an improper translation by half a lattice constant in $x_+$ and $x_-$ direction is also a symmetry, $\{{\rm IC}_{2z}|1/2,-1/2,0\}$.
This operation corresponds to a glide-mirror symmetry and can be expressed as $i \sigma_1\otimes\tau_2$ ($i$ is included to satisfy double group algebra, i.e. ${\rm IC}^2_{2z}=\overline{E}$). As the system is lattice periodic, the operations represent the generators of a factor group $\mathcal{G}/\mathcal{T}$, where $\mathcal{G}$ denotes the full space group and $\mathcal{T}$ the group of pure lattice translations which is an Abelian normal subgroup of $\mathcal{G}$. It follows that there has to be a third nontrivial symmetry element $\left\{{\rm KIC}_{2z}|1/2,-1/2,1/2\right\}$, given by the composition of $\left\{{\rm K}|0,0,1/2\right\}$ and $\left\{{\rm IC}_{2z}|1/2,-1/2,0\right\}$.
This operation corresponds to the map $|\phi\rangle\rightarrow |\pi-\phi\rangle$, realized by the permutations $a_1\leftrightarrow b_1$ and $a_2\leftrightarrow b_2$, which is represented as $-i \sigma_3\otimes\tau_2$ in the basis defined by $\psi_{\bf k}$.
Furthermore, since spin is involved, each of the elements comes with a respective double group partner. The factor group itself is isomorphic to the magnetic double group
$\mathcal{G}/\mathcal{T} \simeq \mathcal{C}_{\text{S}} \oplus \left\{{\rm K}|0,0,1/2\right\}\circ \mathcal{C}_{\text{S}}$,
where $\oplus$ denotes the set sum and $\mathcal{C}_{\text{S}}=\left\{E,\overline{E},IC_{2z},\overline{IC}_{2z}\right\}$.

In the double exchange model for the twisted phase, time-reversal symmetry is broken. Instead the antiunitary symmetries $\left\{{\rm K}|0,0,1/2\right\}$ and $\left\{{\rm KIC}_{2z}|1/2,-1/2,1/2\right\}$ are present. Furthermore, $\left\{{\rm K}|0,0,1/2\right\}^2 = 1$ and $\left\{{\rm KIC}_{2z}|1/2,-1/2,1/2\right\}^2=-1$, which indicates that $\left\{{\rm KIC}_{2z}|1/2,-1/2,1/2\right\}$ can be regarded as an effective time-reversal operation~\cite{SM}. However, as the spin degree of freedom is frozen into the lattice, this effective time-reversal operation acts in a pseudo-spin space spanned by the pseudo-spinor $\psi_{\bf k}$. Another example of effective time-reversal symmetry, with similar algebraic properties was discussed in Ref. \onlinecite{BenW-2017}, in terms of the antiferromagnetic time-reversal symmetry. Additionally, $\left\{{\rm IC}_{2z}|1/2,-1/2,0\right\}$ is a unitary glide-mirror.

A combination of effective time-reversal and the unitary glide-mirror protects the line nodes on the BZ boundary as shown subsequently. We first notice that the product operation acts on the spatial coordinates $(x_+,x_-,z)$ as
\begin{equation}
{\rm KIC}_{2z}\star{\rm IC}_{2z}:
(x_+,x_-,z) \rightarrow (x_++{\tilde a},x_--{\tilde a},z+a),
\end{equation}
while
\begin{equation}
{\rm IC}_{2z}\star{\rm KIC}_{2z}:
(x_+,x_-,z)\rightarrow (x_++{\tilde a},x_--{\tilde a},z-a).
\end{equation}
Therefore,
\begin{equation}\label{eq:commutation}
{\rm KIC}_{2z}\star{\rm IC}_{2z}\\
=e^{2ik_z a}{\rm IC}_{2z}\star{\rm KIC}_{2z},
\end{equation}
implying that these two operators commute and anticommute, respectively at $k_z=0$ and at the BZ boundary plane, $k_z=\pm\pi/2a$. We  have removed the explicit reference to partial translations for notational clarity. Since the unitary operator ${\rm IC}_{2z}^2=-1$, its eigenvalues are $g_\pm=\pm i$. The antiunitary operator  ${\rm KIC}_{2z}^2=-1$, and transforms the momentum as $(k_+,k_-,k_z)\rightarrow (-k_+,-k_-,k_z)$. Therefore each band is Kramers degenerate at ${\bf k}^0=(k_+^0,k_-^0,k_z)$, with  $k_+^0=\pm\pi/{\tilde a}$ and $k_-^0=\pm\pi/{\tilde a}$, representing the surface of the BZ. In addition, on this surface we now take a line $k_z=k_z^0$, with $k_z^0=0$ ($k_z^0=\pm \pi/2a$) and denote the line by ${\ell}$.
Consider a Bloch state on ${\ell}$,   $|\Psi_{\bf k}\rangle$, such that ${\rm IC}_{2z}|\Psi_{\bf k}\rangle=g_+|\Psi_{\bf k}\rangle$. Then,
\begin{align}\label{eq:protection}
&{\rm IC}_{2z}[{\rm KIC}_{2z}|\Psi_{\bf k}\rangle]=e^{2ik_z a}{\rm KIC}_{2z}[{\rm IC}_{2z}|\Psi_{\bf k}\rangle]\nonumber\\
&=e^{2ik_z a}g_-[{\rm KIC}_{2z}|\Psi_{\bf k}\rangle].
\end{align}
Therefore, for the line ${\ell}$ in the middle of the BZ, $k_z^0=0$, the Kramers partners of bands at the same momentum have opposite eigenvalues of the unitary operator and thus anticross. On the other hand, for the line $\ell$ that lies on the BZ boundary,  the eigenvalues of the Kramers partners are the same and the bands can cross along this line. Most importantly, as we just showed,  the crossing is protected by the combination of these two symmetries, i.e. it cannot be removed as long as these symmetries are operative.

Having established the underlying symmetries for the twisted phase, it is possible to construct a general lattice tight binding Hamiltonian $H({\bf k},\phi)$, given by
\begin{equation}\label{eq:symm-cons-Ham}
H({\bf k},\phi)=\sum_{j=1}^{4}\sum_{\alpha=e,o}\sum_{\mu\nu}f^{(j)}_{\mu\nu}({\bf k})F_{\mu\nu}^{j,\alpha}(\phi)\Sigma^{j,\alpha}_{\mu\nu}.
\end{equation}
Here, $F_{\mu\nu}^{j,e,o}(\phi)$ are respectively even and odd functions of $\phi$, index $j$ labels the four combinations of the parities  of the function $f^{(j)}$ under the change of sign of the $x-y$ and the $z$ components of the momentum, and $\Sigma_{\mu\nu}$ are matrices allowed by the symmetries ~\cite{SM}.

\section{Symmetry breaking perturbations}
\begin{figure}[b!]
\centering
\includegraphics[width=0.4\textwidth]{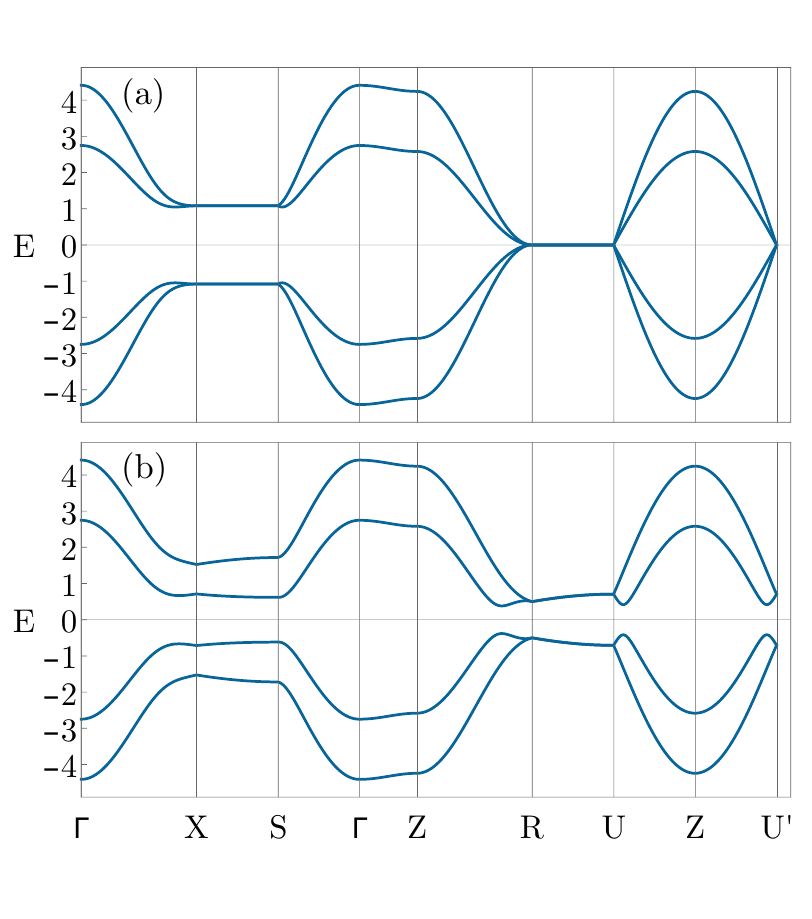}
\caption{\label{fig_pert}Lifting the Hund nodal line via symmetry breaking perturbations. (a) A perturbation introduced by slightly tilting the spins in the $z$-direction preserves Hund NLSM state. (b) A mass term respecting only ${\rm KIC_{2z}}$ symmetry splits the four-fold line node, yielding two pairs of line nodes symmetrically split about $E=0$. A similar effect is obtained when the spins are slightly tilted in the $z$ direction but staggered in the $x-y$ plane.}
\end{figure}
In the following, we consider alternations in the spin structure, which can break the symmetries protecting Hund NLSM. First, we introduce a slight tilt of the magnetic moments in $z$-direction with a staggering between the two layers in the unit cell, breaking ${\rm IC_{2z}}$ and ${\rm K}$, while the ${\rm KIC_{2z}}$ symmetry is kept intact~\cite{SM}.

For a small tilting angle, this perturbation vanishes on the BZ boundary plane $k_x\pm k_y=\pm \pi$,
and consequently preserves the degeneracy of the line node, as shown in Fig.~\ref{fig_pert}(a). Otherwise, it splits the twofold degeneracy of the bands in the $k_z=0$ and $k_z=\pi/2$ planes, since it breaks ${\rm IC_{2z}}$ symmetry.
Furthermore, a perturbation that preserves ${\rm KIC_{2z}}$ but breaks  both ${\rm IC_{2z}}$ and ${\rm K}$ symmetries in contrast gives rise to a mass term which is non-vanishing along the edge of the BZ boundary at $k_z=\pi/2$~\cite{SM}. The two Kramers  pairs of nodal lines are then symmetrically split about the zero energy, while retaining Kramers degeneracy [Fig.~\ref{fig_pert}(b)].

To further illustrate the symmetry protection of the line node, we  consider slightly twisted spins in $z$-direction staggering between $a$- and $b$-sites~\cite{SM}. The corresponding perturbation breaks both ${\rm IC_{2z}}$ and ${\rm KIC2_2}$ symmetries while only the ${\rm K}$ symmetry is kept. Similarly to the previous case, this perturbation leads to a fully gapped band structure (Fig.~\ref{fig_pert}(b) and Sec. S2 of the SM~\cite{SM}).

\section{Surface states}
\begin{figure}[t!]
 \centering
 \includegraphics[width=0.49\textwidth]{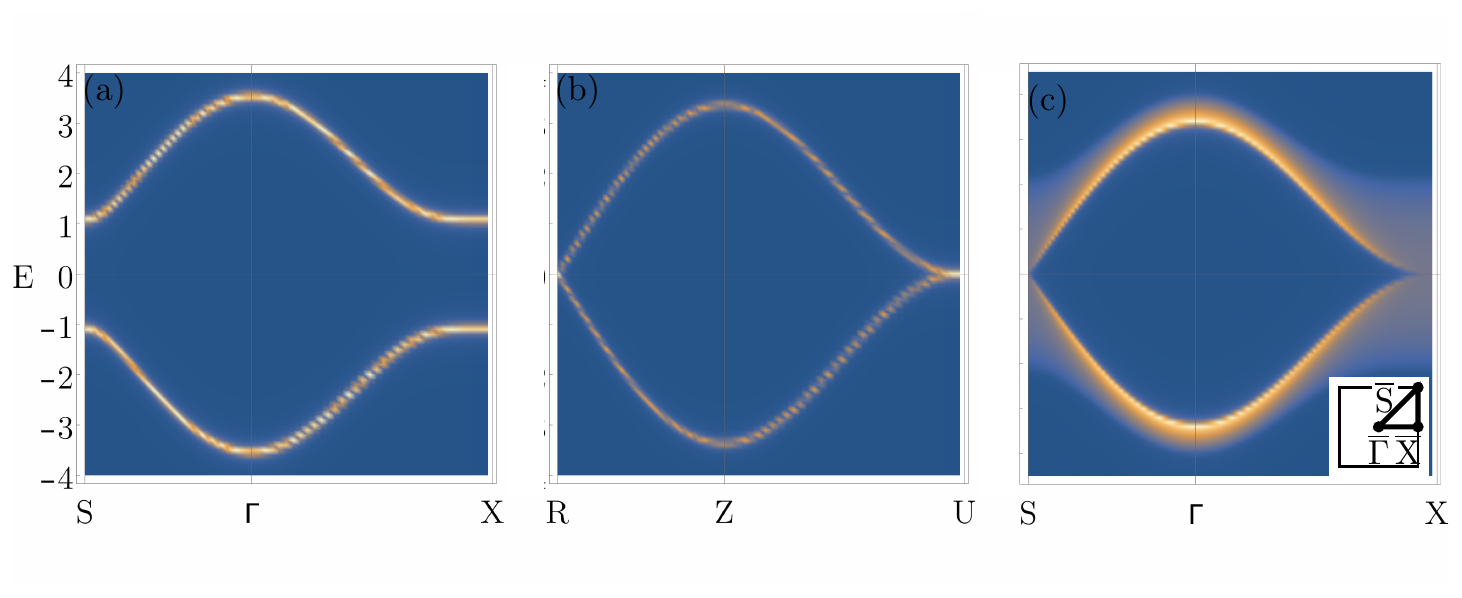}
 \caption{\label{fig_surface} Correspondence between bulk and surface band structure through the calculated local density of states. (a) Bulk band structure in the $k_z=0$ plane along the $S-\Gamma-X$ high symmetry line (not cutting the nodal line); (b) Bulk band structure in the $k_z=\pi/2$ plane along the $R-Z-U$ high symmetry line (cutting the nodal line); (c) Surface band structure for a cut through the crystallographic $(001)$ plane along the high symmetry ${\bar S}-{\bar\Gamma}-{\bar X}$ directions (inset). The local density of states scales with brightness.}
 \end{figure}

To illustrate the topological nature of the model we have calculated the surface electronic structure for a cut along the (001) surface ~\cite{SM}, by means of the ${\bf k}$-dependent local density of states or Bloch spectral function projected on the surface layer. Fig. \ref{fig_surface} compares the energy dispersion for the bulk and the surface. The surface band structure features a dispersing drumhead state with maximal localization at $\overline{\mathrm{\Gamma}}$ (the center of the surface BZ) ending with completely delocalized states at the edge of the surface BZ which are connected to the bulk nodal lines. This in turn demonstrates the bulk-surface correspondence for the Hund NLSM.

\section{Discussion and Conclusions}
We show that nodal line semimetals protected  by non-symmorphic unitary and anti-unitary symmetries can emerge out of a strong Hund's coupling  between the localized spins forming a noncollinear magnetic phase and the itinerant electrons. This situation is likely to exist in systems described by the so called double exchange model, like the Colossal Magneto Resistance manganite oxides~\cite{AH55,CVM99}. In that respect, prominent candidates include La$_{2-2x}$Sr$_{1+2x}$Mn$_2$O$_7$ for $x=0.58$ reported to be metallic with a magnetic state consistent with a twisted in-plane spin configuration~\cite{Badica2004}, as well as Nd$_{1-x}$Sr$_x$MnO$_3$ for $x=0.5-0.6$~\cite{Kajimoto1999}, and Ca$_{1-x}$Sm$_x$MnO$_3$, for $x<0.12$~\cite{Maignan1998} both also metallic  with a possible non-collinear spin state. In addition, the rich phase diagram of the double exchange materials allows for electronic phase separation~\cite{Yetal98,AG98}, and the coexistence of phases with different topologies. Nontrivial interface states are expected to emerge. \change{Response to magnetic topological defects, such as vortices and skyrmions (for spin textures with an out-of-plane spin component), through binding of special mid-gap states may further distinguish magnetic topological states in comparison with their non-magnetic counterparts~\cite{Ran2009,Juricic2012}}. Our findings therefore open up a route for studying the emergence of exotic states of matter in materials where localized spins and itinerant electrons are strongly coupled.

This should motivate further studies of the experimental imprints of this class of topological metals, through for instance, the drumhead surface states, in the tunneling experiments and transport. Furthermore,  the bulk magnetotransport signatures should be observable, as a direct manifestation  of the recently proposed parity anomaly for NLSMs~\cite{rui2018,burkov2018}. Finally, the role of interactions~\cite{RGJ} and different types of disorder~\cite{RSJ} in this context are still open problems.

\section{Acknowledgement}\label{acknowledgements}
VJ is grateful to Leslie Schoop for useful discussions. RMG acknowledges support from the Swedish Research Council Grant No.~638-2013-9243, the Knut and Alice Wallenberg Foundation, and the European Research Council under the European Union’s Seventh Framework Program (FP/2207-2013)/ERC Grant Agreement No.~DM-321031.

\bibliography{references_v2}
\end{document}